\documentstyle[aps,prl,twocolumn,floats,epsfig]{revtex}

\begin{document}
\draft
\wideabs{

\title{Cold collisions between atoms in optical lattices}

\author{J. Piilo,$^1$ K.-A. Suominen,$^{1,2,3}$ and K. Berg-S{\o}rensen$^{4}$}

\address{$^1$Helsinki Institute of Physics, PL 9, FIN-00014 Helsingin
yliopisto,
Finland}

\address{$^2$Department of Applied Physics, University of Turku,
FIN-20014 Turun yliopisto, Finland}

\address{$^3${\O}rsted Laboratory, Universitetsparken 5, DK-2100
Copenhagen {\O},
Denmark}

\address{$^4$Nordita, Blegdamsvej 17, DK-2100 Copenhagen {\O}, Denmark}

\maketitle

\begin{abstract}
We have simulated binary collisions between atoms in 
optical lattices during Sisyphus
cooling. Our Monte Carlo Wave Function simulations show that the collisions
selectively accelerate mainly the hotter atoms in the thermal 
ensemble, and thus
affect the steady state which one would normally expect to reach in Sisyphus
cooling without collisions.
\end{abstract}

}

\narrowtext

Neutral atoms can be cooled and trapped in light-induced optical
lattices~\cite{Jessen96}. By controlling the laser light one can adjust the
properties of the lattices in order to study e.g.~the quantum nature of atomic
motion in a periodic structure~\cite{Castin91}, including the analogues to the
behavior of electrons in periodic solid state lattices~\cite{Birkl95}. Ideas
regarding the possibility to use optical lattices in atom optics and quantum
computation have also emerged
recently~\cite{Adams94,Brennen99,Jaksch99,Hemmerich99}. In 
experiments the trapped
gas density is typically very low, providing for the moment at best a filling
ratio of $10\%$ for the near red detuned lattices~\cite{Jessen96}. Thus it is
{\it a priori} a good approximation to ignore that the atoms interact with each
other. In magneto-optical traps for cold atoms the inelastic 
collisions limit the
numbers and temperatures achievable for the atomic gas as densities increase to
about 10$^{11}$ atoms/cm$^3$~\cite{Weiner99}. By using Bose-Einstein 
condensates
or combining lattices and other types of optical traps it is becoming 
possible to
obtain filling ratios close to unity and even higher~\cite{Jaksch98}. We have
considered mainly the case where the filling ratio is about $25\%$, 
but our results
can be qualitatively interpolated for smaller ratios. Applications 
such as quantum
computing require atoms to interact in order to perform quantum logical
operations~\cite{Brennen99,Jaksch99}.

Controlled interaction studies in optical lattices could be performed e.g.~by
superimposing two optical lattices, which can be moved in respect to each other
\cite{Jaksch99}. This, however, does not answer the question of what 
happens in a
basic lattice configuration when the filling ratio increases, especially when
inelastic collisions interfere with the cooling process and 
localization of atoms
at lattice sites. For low densities the atom cloud reaches a thermal 
equilibrium
state, and based on the studies in magneto-optical traps one would expect that
inelastic collisions increase the temperature of this equilibrium state as the
gas density increases~\cite{Holland94}. We have performed Monte Carlo Wave
Function (MCWF) simulations of two atoms in a lattice. They show that (for the
parameters of our study) an equilibrium is not easily obtained. Instead, the
hotter atoms are selectively accelerated, and, especially in 2D lattices, are
likely to leave the lattice. Thus in densely populated lattices  
Sisyphus cooling
could be assisted by an evaporation process: interactions eject the 
hotter atoms
whereas the remaining atoms thermalize via Sisyphus cooling (in contrast to the
collisional thermalization in evaporative cooling in magnetic traps).

In a collision two cold atoms get close enough to form a long-range
quasimolecule~\cite{Weiner99}. Compared to single atoms, the quasimolecule
interacts differently with the surrounding laser light, and this interaction
depends on the interatomic distance. Previously the atomic interactions in
lattices have been modelled by assuming fixed positions for 
both atoms and
calculating how the atomic energy levels are shifted by the
interaction~\cite{Goldstein96,Boisseau96,Guzman98,Menotti99}. Such 
static models
ignore the dynamical nature of the inelastic collisions. But to allow 
the atoms to
move makes the problem complicated and computationally tedious. We 
present in this
Rapid Communication a study of collisions in a lattice between {\it 
moving} atoms.
Once the dynamical processes are understood, they can be used as input for
macroscopic theories. This approach leaves out many other aspects of 
the problem,
such as reabsorption of scattered photons, which is another mechanism that
strongly limits the densities in magneto-optical traps. Thus our results do not
necessarily reflect the complete situation in optical lattices, but we believe
they demonstrate the effect of collisions on Sisyphus cooling. A more complete
study is simply beyond the modern computational resources.

The distribution of atoms in an optical lattice depends on the choice of laser
field configuration and the atomic level structure. The laser field should
have a spatially changing polarization, and the atom needs at least two Zeeman
sublevels in the lower energy state, and a different angular momentum 
in the upper
energy state.  The interaction between the laser field and an atom 
gives rise to
periodic light-induced potentials for atoms in the Zeeman states of 
their internal
ground states. A single atom moving in such a lattice will undergo Sisyphus
cooling because of optical pumping from one ground state to another, 
in a manner
that favors the reduction of kinetic energy between the rapid optical pumping
cycles~\cite{Dalibard89}. This cooling effect takes place rapidly in lattices
created by lasers which are tuned only a few linewidths below the atomic
transition. After cooling, the atoms 
are to a large
extent localized in these potential wells.

We have chosen as a basis for our studies the simplest atomic transition for a
red-detuned laser field, i.e., a system with a lower state angular momentum
$J_g=1/2$ and an upper state angular momentum $J_e=3/2$. We denote the first
state as the ground state $|g_{\pm 1/2}\rangle$, the index referring to the
quantum number $m$ for the eigenvalue of the $z$ projection of the 
angular momentum
operator, $J_z$. Similarly, we denote the second state as the excited state,
with eigenstates $|e_{\pm 3/2}\rangle$ and $|e_{\pm 1/2}\rangle$. The resonance
frequency of the transition is $\omega_0$. In the numerical calculations
we have used the atomic properties of Cs.

The laser field has periodicity in one dimension, and consists of two linearly
polarized counter-propagating beams, with orthogonal linear polarization and
frequency $\omega$. For this configuration, the combined laser field is
\begin{equation}
     {\bf E}(z,t)={\cal E}_0 ({\bf e}_x e^{ikz} - i {\bf e}_y
     e^{-ikz})e^{-i\omega t} +
     c.c., \label{eq:Efield}
\end{equation}
where ${\cal E}_0$ is the amplitude and $k$ is the wavenumber.

When the interactions become important, the atomic cloud is still relatively
dilute so that only two atoms at a time are involved, and the dipole-dipole
interaction (DDI) dominates the process. We calculate the two-atom DDI
potentials following the procedure described in Appendix A of 
Ref.~\cite{Lenz93}.
We consider two atoms interacting with the laser field, coupled to a reservoir,
namely the vacuum electric field. The system Hamiltonian reads (after rotating
wave approximation)
\begin{equation}
     H_s=\sum_{\alpha=1,2} \frac{p_{\alpha}^{2}}{2M} - \hbar \delta P_{e,\alpha}
     + V, \label{eq:Hsys}
\end{equation}
where the sum over $\alpha$ is over the two atoms, $\delta$ is the detuning
$\delta=\omega - \omega_0$, $M=133$ a.u. is the Cs atom mass, and 
$P_{e,\alpha} =
\sum_{m=-3/2}^{3/2} |e_m \rangle_{\alpha}~_{\alpha}\langle e_m|$. The 
potential $V$
gives the interaction with the laser field. The strength of this interaction is
given by the Rabi frequency $\Omega= 2d{\cal E}_0/\hbar$ where $d$ is 
the dipole
moment of the transition.

The system interacts with the reservoir through a dipole coupling between the
atoms and the vacuum modes. As we want an expression for the DDI potential, we
concentrate our effort on the calculations leading to an expression like
Eq.~(25) of Ref.~\cite{Lenz93}, which originates from the commutator
between the system density operator, $\rho$, and the DDI potential, $V_{dip}$.
Also, we concentrate on spontaneous terms, i.e., terms with vanishing average
photon number. Let us introduce the operators
\begin{equation}
     S_{+,q}^{\alpha}=\sum_{m=-1/2}^{m=1/2} CG_{m}^{q}
     |e_{m+q}\rangle_\alpha ~_\alpha\langle g_m|,
     \label{eq:S+q-alpha}
\end{equation}
where $CG_{m}^{q}$ are the appropriate Clebsch-Gordan coefficients and $q$
is the polarization label in spherical basis. Furthermore, we use a 
description in
terms of a center of mass coordinate $Z$ and a relative coordinate 
${\bf r}={\bf
r}_2-{\bf r}_1$ (with coordinate along the quantization axis 
$z=z_2-z_1$).  With
these coordinates, the interaction potential with the laser field reads
\begin{eqnarray}
     V&=&-i\frac{\hbar \Omega}{\sqrt{2}}\sin kZ \cos k\frac{z}{2} S_{+,+}
     +i \frac{\hbar \Omega}{\sqrt{2}} \cos kZ \sin k\frac{z}{2} \Delta S_{+,+}
     \nonumber \\
     &&+\frac{\hbar \Omega}{\sqrt{2}} \cos kZ \cos k\frac{z}{2} S_{+,-}
     +\frac{\hbar \Omega}{\sqrt{2}} \sin kZ \sin k\frac{z}{2} \Delta S_{+,-}
     \nonumber \\
     &&+h.c., \label{eq:V}
\end{eqnarray}
where $S_{+,q} = S_{+,q}^{1} +S_{+,q}^{2}$, and $\Delta S_{+,q}= S_{+,q}^{1}-
S_{+,q}^{2}$.

In order to calculate the DDI term, we look at the Hamiltonian part 
of the damping
terms in the equation of motion for the system density operator $\rho$. After
manipulations similar to those presented in Appendix A of 
Ref.~\cite{Lenz93}, and
using arguments from Ref.~\cite{Berman97} to evaluate integrals of Bessel
functions multiplied with principal value functions, we find the DDI potentials
between the two atoms. In the following we look only at atoms on the 
axis of the
laser field, i.e., a one-dimensional situation, and in this case, the 
DDI potential
reduces to
\begin{eqnarray}
     V_{dip}^{axis}&=&\frac{3}{8}\hbar\Gamma \left\{ \frac{1}{3}
     \frac{\cos q_0 r}{q_0 r}
     +2\left[ \frac{\sin q_0 r}{(q_0 r)^2} + \frac{\cos q_0 r}{(q_0 r)^3}
     \right] \right\} \times \nonumber \\
     &&~~~\left( {\cal S}_{++}{\cal S}_{-+} + {\cal S}_{+-}{\cal S}_{--}
     - 2{\cal S}_{+0}{\cal S}_{-0} \right).
     \label{eq:Vdip-axis}
\end{eqnarray}
Here, $\Gamma$ is the atomic linewidth, $q_0$ is the resonant wavenumber
$q_0=\omega_0/c$, and
\begin{equation}
     {\cal S}_{+q} {\cal S}_{-q'} \equiv
     \left( S_{+,q}^{1}S_{-,q'}^{2} + S_{+,q}^{2} S_{-,q'}^{1}
     \right).
  \label{eq:S+qS-q}
\end{equation}

Numerical simulation of the motion of atoms in the lattice field in 
one dimension
only, using the MCWF method~\cite{Dalibard92}, is computationally very
demanding~\cite{cpu}. In order to perform two-atom studies, which 
require even in
one dimension at least two translational degrees of freedom, we have fixed one
atom in position, and let the other one move freely. This fixes the relation
between the lattice coordinates and the relative interatomic 
coordinate. Thus an
inelastic collision will not change the kinetic energy for both 
atoms, but we use
the relative kinetic energy as an estimate for the kinetic energy 
change per atom.
(We express energy and momentum in recoil units: $E_r=\hbar^2k^2/2 M$ and
$p_r=\hbar k$ respectively).

We have formulated the problem in the two-atom basis, which leads to a system
of 36 internal states. In studies for magneto-optical traps one tends to use
the molecular frame, where the atom-atom interactions have been included to the
molecular potential structure~\cite{Julienne91}. However, the quantum jump
processes needed for the Monte Carlo method are easier to describe in 
the atomic
basis. One aspect of the simulations is that we do not use the adiabatic
elimination of the excited states~\cite{Petsas99}, which is typically 
employed in
order to simplify the equations for atomic motion. For simplicity we neglect
Doppler cooling (as Sisyphus cooling takes us below the Doppler 
limit, we expect
it to be the dominant process). In the molecular frame the system of two
interacting atoms is excited resonantly to a molecular state with an attractive
interatomic potential~\cite{Julienne91}. This leads to the acceleration of the
relative motion of the atoms, until the process terminates with 
spontaneous decay.
We use these attractive potentials for the verbal description of the 
process but
it must be emphasised that they do not directly appear in the 
two-atom basis. The
kinetic energy change due to the attractive potentials also complicates greatly
the numerical simulations by demanding larger momentum and finer 
spatial grid than
in the single atom Sisyphus cooling simulations~\cite{Numerics}.

We use the laser parameters $\delta= -3\Gamma$ and $\Omega= 
1.5\Gamma$, which give
a lattice modulation depth of $U_{0}=584E_{r}$. These parameters 
correspond to a
lattice where the atoms move from one lattice site to another on a 
timescale that
is comparable to the timescale of a harmonic oscillation within one 
of the lattice
potential wells. In our selected system the atomic interactions are too weak to
really destroy the lattice, so the actual case of interest is the one where the
atoms need to be simultaneously at the same lattice site.

In the MCWF method an approximation for the two-atom steady state 
density matrix
is obtained as an ensemble average of different wave function histories, for
which the spontaneous emission occurs as probabilistic quantum
jumps~\cite{Dalibard92}. These quantum jumps (both atoms in our case 
have six decay
channels) occur according to probabilities weighted by the appropriate
Clebsch-Gordan coefficients of the decay channels. There are various 
ways how to
calculate the results by ensemble averaging. We take the ensemble average of
single history time averages in the steady state time 
domain~\cite{Molmer96}. Thus
we obtain the kinetic energy per atom, and the spatial and momentum probability
distributions for various filling ratios $(\rho_o)$ of the lattice.

A comparison between the number of atoms having gained large kinetic energy via
interactions and the total number of interaction processes show (see
Table~\ref{uo600na}) that basically every collision produces very hot 
atoms in our
chosen parameter range. This leads to evaporation in the optical lattice: those
atoms which are able to move from one well to the other and which have larger
kinetic energy than localized atoms leave the trap. A crucial ingredient in the
interaction process  increasing the kinetic energy by a large amount 
and leading
to evaporation is that a large fraction of the population has to enter the
attractive molecular excited state during the interaction process. 
This fraction
in turn depends on the relative velocity between the interacting 
atoms when they
reach the resonance point for the attractive molecular states. The relative
velocity in turn depends on the lattice depth. In our simulations the 
surroundings
is still favorable so that the relative velocity between atoms is low enough to
keep the excitation probability high when atoms approach each other 
and cross the
molecular resonance point.

The number of attractive molecular states is five
and the resonant excitation to these potentials takes place at different
interatomic distances~\cite{Julienne91}. If the atoms do not get a 
large increase
in kinetic energy at the first resonance they reach, there are still other
resonances left. A comparison with semiclassical (SC) excitation and survival
calculations suggests that the potential which becomes resonant first when the
atoms approach each other has the dominant role in the inelastic collision
process.

When calculating the steady state kinetic energy per atom (Table 
\ref{uo600na}),
we use two critical wavenumbers $k_c$. Wavefunction histories which at some time
point have gained larger total kinetic energy than given by $k_c$ are neglected
in ensemble averaging (considered lost from the lattice). The smallest value of
$k_c$ we use~\cite{kc} is more than two times larger than the semiclassical
critical value $k_c^{sc}$ given in Ref.~\cite{Dalibard89}. The denser 
the lattice
is initially, the larger is the number of interaction processes and the more
effective is the evaporative cooling process. This can be seen in the 
results for
kinetic energy per atom using $k_c=40$ (see Table~\ref{uo600na}). The kinetic
energy decreases when the initial density of the lattice increases. 
Results with
$k_c=70$ include atoms that are lost from the lattice, and the value of 
the kinetic energy is slightly above the sparse lattice 
(non-interacting case) result.

\begin{figure}[ht!]
\centering
\psfig{figure=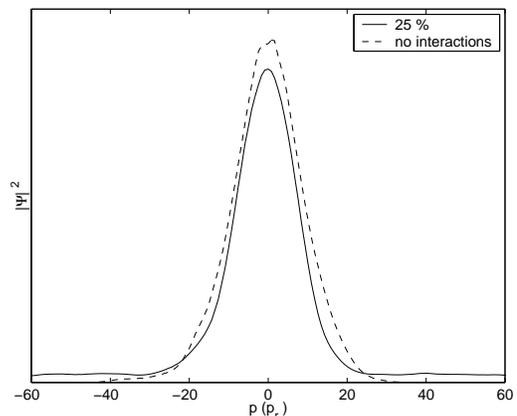,scale=0.4}
\caption[f1]{\label{uo600kd60}
The steady state momentum probability distributions for a densely populated
$(\rho_o=25\%)$ lattice and for the non-interacting atoms case (see 
text). All of
the MC histories are included. Here $\delta = -3.0\Gamma$, and $\Omega =
1.5\Gamma$. }
\end{figure}

The momentum distribution in Fig.~\ref{uo600kd60} shows the effect of the
evaporative cooling process clearly.  Due to the interactions between 
atoms part
of the population has shifted to the region of large $k$ (wings in
Fig.~\ref{uo600kd60}) and does not localize back to the lattice 
because the atoms
are above the recapture range. Thus the central peak of the momentum 
distribution
corresponding to atoms localized at lattice sites has a $13\%$ 
narrower FWHM for
an initially dense lattice compared to the non-interacting case.

We have shown that in high-density, red-detuned (a few linewidths) optical
lattices, atomic interactions could lead to the ejection of the hotter atoms
from the lattice, or at least to a selective heating process
accompanied by a narrowing of the central momentum distribution.
This is because (a)
atoms may move from one lattice site to another even in the steady state for
Sisyphus cooling, and (b) because the molecular interaction is strong enough to
give to each clearly interacting atom pair almost always enough 
energy to escape
from the lattice. Earlier simulations for roughly the same laser (and Cs)
parameters in magneto-optical traps indicate that the dominating 
effect is a clear
broadening of the atomic momentum distribution, i.e., radiative
heating~\cite{Holland94}. In both situations high-momentum atoms are 
produced, but
in optical lattices atoms with higher momentum are strongly 
favoured in the momentum increasing process (in
both cases the fast atoms get involved in more close encounters than the slow
ones, but in lattices working in our selected parameter region this 
fact becomes
enhanced).

J.P. and K.-A.S. acknowledge the Academy of Finland (project 43336), NorFA,
and Nordita for financial support, and the Finnish Center for Scientific
Computing (CSC) for computing resources. J.P. acknowledges support from the
National Graduate School on Modern Optics and Photonics.

\begin{table}
\caption{\label{uo600na}
Escaped atoms and kinetic energies. The number of MC histories which have been
neglected ($N_{out}$) in the ensemble averaging due to escape from 
the lattice and
steady state kinetic energies per atom ($<E_{k}>$)  for various filling ratios
$(\rho_o)$ of the lattice. Two different critical 
wavenumbers $k_c$ have
been used. $N^{int}_{tot}$ gives an estimate for the total number of atom-atom
interaction processes based on single atom MC collision rate calculations by
monitoring the quantum flux at a mean atomic separation given by $\rho_o$.
The total number of MC histories for each simulation is $128$. The 
absolute values
of the standard deviation for the kinetic energies are given in parentheses.
}
\begin{center}
\begin{tabular}{ccccccccc}
   $\displaystyle{\rho_o (\%)}$
& $\displaystyle{N^{int}_{tot}}$
& $\displaystyle{N_{out}}$
& $\displaystyle{N_{out}}$
& $\displaystyle{<E_{k}>}$
& $\displaystyle{<E_{k}>}$
\\
&
& $\displaystyle{k_{c} = 40 }$
& $\displaystyle{k_{c} = 70 }$
& $\displaystyle{k_{c} = 40 }$
& $\displaystyle{k_{c} = 70 }$
&
\\
\hline
   25.0 & 39 &38 & 26 & 61(6) & 110(18) &\\
   20.0 & 25 &27 & 19 & 69(5) & 99(12)  &\\
   16.7 & 19 &19 & 11 & 80(6) & 103(12) &\\
   14.3 & 16 &19 & 12 & 80(7) & 104(12) &\\
no interactions & 0 & 0 & 0& 91(8)& 91(8)& \\
\end{tabular}
\end{center}
\end{table}

\begin{references}
\bibitem{Numerics}
Due to the large kinetic energies, the spatial step size $\Delta z$ has
to be smaller than in single atom simulations since the momentum space size is
$L_k=2 \pi/\Delta z$. We use the split operator Fourier method to solve the
non-Hermitian time-dependent Schr\"odinger equation. The lattice density is
controlled with the position space size $L_z$.
\bibitem{kc}
At the semiclassical value $k_c^{sc}$~\cite{Dalibard89} the cooling 
force has its
maximum value, but Sisyphus cooling is still effective above that point. Our
values of $k_c$ gives a criterion for neglecting energetic histories from the
ensemble (i.e., atoms lost from the lattice).
\end{references}
\end{document}